\documentclass{aa}  
\usepackage{graphicx}
\usepackage{txfonts}
\usepackage{natbib}
\usepackage{booktabs}
\usepackage{tabularx} 
\bibpunct{(}{)}{;}{a}{}{,}   
\begin{document}
   \title{Discovery of a QPO in the X-ray pulsar 1A~1118-615:\\
    correlated spectral and aperiodic variability}


    \author{E. Nespoli
          \inst{1}
        \and
          P. Reig\inst{2,3}
          }

   \offprints{Elisa Nespoli}

   \institute{Observatorio Astron\'omico de la Universidad de Valencia, Calle Catedr\'atico Agust\'in Escardino Benlloch 7, 46980 Paterna, Valencia, Spain\\
       \email{elisa.nespoli@uv.es}
                \and
             Foundation for Research and Technology -- Hellas, IESL, Voutes, 71110 Heraklion, Crete, Greece
             \and
             Physics Department, University of Crete, 710 03 Heraklion, Crete, Greece\\
             }


 
  \abstract
   {}
   {Our goal is to investigate the X-ray timing and spectral variability of the high-mass X-ray binary 1A~1118-615 during a type-II outburst.}
   {We performed \  {a} detailed color, spectral and timing analysis of a giant outburst from 1A~1118-615, using \emph{RXTE} data. }
   {We report the discovery of a variable quasi-periodic oscillation (QPO) in the power spectral density of 1A~1118-615, with a centroid frequency of $\sim$0.08 Hz. The centroid frequency of the QPO correlates with the X-ray flux, as expected according to the most accredited models for QPO production. For energies above $\sim$4 keV, the QPO $rms$ variability decreases as the energy increases. 
Pulse profiles display energy dependence, with a two-peak profile at lower energies, and a single peak at higher energies.    
 From spectral analysis, we confirm the presence of a cyclotron absorption feature at $\sim$60 keV, the highest value measured for an X-ray pulsar. We find that the spectral parameters (photon index, cutoff energy, iron fluorescence line strength) display a marked dependence with flux. We detect two \  {different levels of neutral hydrogen column density}, possibly due to the Be companion activity. We report for the first time a correlation between the timing and spectral parameters in an X-ray pulsar. All the correlations found between spectral/timing parameters and X-ray flux are \  {present} up to a flux of $\sim$6$\times$10$^{-9}$ erg cm$^{-2}$ s$^{-1}$, when a saturation level is reached. We propose that the saturation observed corresponds to the minimum extent of the neutron star magnetosphere. We estimate the magnetic field of the neutron star from two independent ways, using results from spectral (cyclotron line energy) and timing (QPO frequency) analysis, obtaining consistent values, of $\sim$$7-8 \times10^{12}$ G. Results from the comprehensive spectral and timing analysis are discussed in comparison with other X-ray pulsars.}
   {}


\titlerunning{Discovery of a QPO in 1A~1118-615: correlated spectral and aperiodic variability}
   \maketitle
%

\section{Introduction}
Be/X-ray binary systems (Be/XRBs) constitute a subclass of neutron-star high-mass X-ray binaries (HMXBs) in which a neutron  star orbits a Be star. 
In these systems, the Be-stars are deep inside their Roche lobes, as is indicated by their generally long orbital periods and by the absence of X-ray eclipses and of ellipsoidal light variations. The X-ray emission from Be/XRBs tends to be extremely variable, ranging from quiescence, to periodic, short  (Type-I) outbursts separated by the orbital period  ($L_\mathrm{X} \sim 10^{36}-10^{37}$ erg s$^{-1}$), to giant (Type-II) transient outbursts lasting weeks to months ($L_\mathrm{X} \gtrsim 10^{37}$ erg s$^{-1}$). Be/XRBs show hard X-ray spectra which, in combination with the regular X-ray pulsations, indicate that the compact object must be a strongly magnetized neutron star \citep[see][for a review]{neg07}.\\

1A~1118-615 is an accretion-powered X-ray pulsar which was discovered in December 1974 by the Ariel V satellite \citep{eyl75} during an observation of Cen X-3. The same data revealed pulsations with a period of 405$\pm$0.6 s \citep{ive75}. The X-ray spectrum was fitted by a highly-absorbed ($N_\mathrm{H} \approx 6 \times10^{22}$ cm$^{-2}$) power law with photon index $\sim 1$. The remarkable measured absorption indicated that the  circumstellar environment played a significant role in absorbing the X-ray emission \citep{ive75}. In 1979 and 1985, the system was observed by Einstein and EXOSAT satellites, respectively \citep{mot88}. Only weak signal was detected in both occasions, showing that low-level accretion was occurring. No spectral or temporal analysis could be performed.\\

The second giant outburst from 1A 1118-615 since its discovery was detected in January 1992 after a long period of quiescence.
A multiwavelength study was performed by \citet{coe94}. They included BATSE data in the energy range 20--100 keV, supported by optical, IR and UV observations. The pulsed X-ray spectrum was fitted by a single-temperature, optically-thin, thermal bremsstrahlung model with a temperature of 15 keV. A spin-up of 0.016 s day$^{-1}$ was observed during the decay of the outburst. Lower frequency observations showed very strong H$\alpha$ emission (EW$_{\mathrm{H}\alpha}$= -80\AA), and high IR excess from the disk of the companion star ($E(J-K)$=0.47), indicating a very large circumstellar disk \citep{coe94}. \\

In contrast to what had been previously found by \citet{mot88}, pulsations from 1A 1118-615 were detected by \citet{rut07} at low luminosity ($L_\mathrm{X} \lesssim 10^{35}$ erg s$^{-1}$), making it the third HMXB to exhibit pulsed emission at low luminosities, after A0535+26 \citep{neg00} and 4U 1145--619 \citep{mer87}. After 17 years of quiescence, a third major outburst was observed at the
beginning of 2009. Using data from the Rossi X-ray Timing Explorer ({\it RXTE}), \citet{dor10} measured the broad-band (3.5--120 keV) spectrum of 1A 1118-615 for the first time. From observations corresponding to the peak of the outburst, they detected a cyclotron resonant scattering feature (CRSF) at $\sim$55 keV and the presence of an iron emission line at 6.4 keV.\\

1A 1118-615 is associated with the O9.5 III-Ve star He3--640, with an estimated distance of 5$\pm2$ kpc \citep{jan81}. \  {\citet{vil99} detected strong spectral variability of He3--640, with a correlation between the hard X-ray flux and the H$\alpha$ equivalent width. They proposed the presence of extended, expanding material around the Be star that sometimes becomes dense enough in the vicinity of the neutron star to form an accretion disk; this increases the mass accretion rate, hence the X-ray flux.} {The formation of an accretion disk during Type II outbursts is still under debate. However, the discovery of quasi-periodic oscillations in some systems \citep[see][and references therein]{mar10} would support this scenario. The presence of an accretion disk also helps explain the large and steady spin-up rates seen during the giant outbursts, which are difficult to account for by means of direct accretion. More recently, \citet{lin10} performed a {\it Swift} spectral analysis of 1A 1118-615 during the 2009 outburst and a flaring event two months later and found evidence for emission from an accretion disk during the flare event, suggesting that the polar cap and the accretion disk emission might co-exist after an outburst.}\\

The orbital period of the binary is still unknown, but it can be estimated in the range of 400--800 days from the pulse period/orbital period diagram of \citet{cor86}. \\

In this paper, we present a detailed spectral and temporal study of the third major outburst of the system using {\it RXTE} data.  Unlike previous work, we focus on the aperiodic variability of the source, reporting the discovery of a $\sim$0.08 Hz QPO, we study the evolution of the spectral and timing parameters throughout the entire outburst, and perform pulse-phase spectroscopy. Section 2 describes the observations and gives information on the data reduction. In Sect. 3 we report the results obtained. In Sect. 4 we interpret our findings and compare the variability observed in 1A 1118-615 with that reported in other HMXBs.

\section{Observations and data analysis}

The increasing flux from 1A 1118-615 at the beginning of January 2009 triggered a public \emph{RXTE}  Target of Opportunity (ToO) campaign of observations. Table~\ref{tab:obslog} shows the observation log. \\

 \begin{table}[!h]
      \caption{\small{Journal of RXTE observations.}}
      \begin{center}
        \centering
            \begin{tabular}{cccc}
             \hline
             \hline
             \noalign{\smallskip}
                N. of           & Proposal  & MJD  & On-source     \\
                pointings &  ID &            range &  time (ks) \\
             \noalign{\smallskip}
              \hline
             \noalign{\smallskip}
             1  & 94412  & 54838.4                      & 1.30 \\ 
              25   & 94032  &  54841.2-- 54865.0     & 8.90 \\                                               
         \noalign{\smallskip}
            \hline
\end{tabular}  
         \end{center}
   \label{tab:obslog}
  \end{table}

In this work, data from both the Proportional Counter Array (PCA) and the High Energy Timing Experiment (HEXTE) onboard \emph{RXTE} \citep{bra93} were employed. The PCA consists of
 five Proportional Counter Units (PCUs) with a total collecting area of  $\sim$6250 cm$^{2}$, operates in the 2--60 keV range and has a nominal energy resolution of 18\% at 6 keV. The HEXTE is constituted by 2 clusters of 4 NaI/CsI scintillation counters, with a total collecting area of 2 $\times$ 800 cm$^{2}$, sensitive in the 15--250 keV band with a nominal energy resolution of 15\% at 60 keV. Both instruments have a maximum time resolution of $\sim$1$\mu s$\\
 
Data reduction was performed using HEASOFT version 6.9. 
An energy spectrum was obtained for each one of the 26 observations, after filtering out unsuitable data according to the recommended criteria\footnote{Among which, elevation ftrom the Earth greater than 10$^\circ$ and pointing offset lower than 0.02$^\circ$; see PCA digest at http://heasarc.gsfc.nasa.gov/docs/xte/pca$\_$news.html}, employing Standard 2 mode data from the PCA (PCU2 only) and Standard (archive) mode from the HEXTE. At the time of the observations, the HEXTE Cluster A had ceased its modulation between on-source and off-source positions, while Cluster B was still rocking: for this reason we only employed Cluster B data, excluding detector 2 since its inability to measure spectral information. The PCA and HEXTE spectra were extracted, background subtracted and deadtime corrected. For the PCA, the 3--30 keV energy range was retained, while the HEXTE provided a partially overlapping extension from 25 to 100 keV. For each observation, the two resulting spectra were simultaneously fitted with XSPEC v.~12.6.0 \citep{arn96}. During the fitting, a systematic error of 0.6\% was added to the spectra.\\

Power spectral density (PSD) was computed using PCA Good$\_$Xenon data. \  {In one case (observation 94412-01-01-00, the first of the data set) Event mode data were employed.} 
We first extracted, for each observation, a light curve in the energy range 2--20 keV (channels 0--49) with a time resolution of $2^{-5}$ s. The light curve was then divided into 128-s segments and a Fast Fourier Transform was computed for each segment. The final PSD was computed as the average of all the power spectra obtained for each segment. These averaged power spectra were logarithmically rebinned in frequency and corrected for dead time effects according to the prescriptions given in \citet{now99}. Power spectra were normalized such that the integral over the PSD is equal to the squared fractional $rms$ amplitude, according to the so-called $rms$-normalization \citep{bel90,miy91}.

   \begin{figure}
   \includegraphics[width=9.8cm]{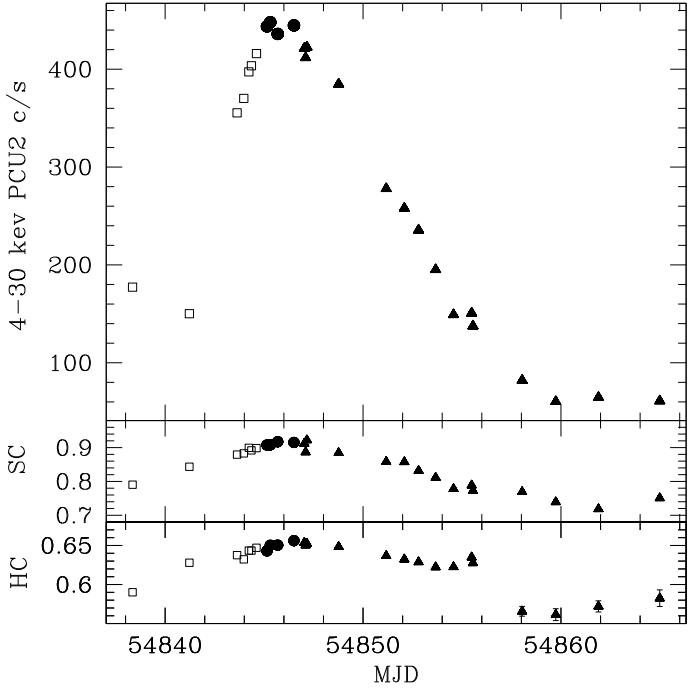}   
  \caption{\small{PCA light curve and color behavior during the 2009 outburst.}}
      \label{fig:lc}
   \end{figure}

\section{Results}

1A 1118-615 went into outburst at the beginning of 2009. A flux clearly above the background was detected on MJD 54840. The total duration of the outburst was 27 days. A maximum \  {3--30 keV} X-ray luminosity of $2.8 \times 10^{37}$ erg s$^{-1}$ was measured at MJD $\sim 54845.3$, assuming a distance of 5 kpc.\\
 
Figure~\ref{fig:lc} shows the PCA light curve and color behavior during the outburst, as directly derived from PCU2 count-rate. Each point in the diagrams corresponds to a \emph{RXTE} pointing. The color definition is the following: soft color (SC): 7--10 keV / 4--7 keV; hard color (HC): 15--30 keV / 10--15 keV. Different symbols mark the different phases of the outburst, the rising (open squares), the peak (filled circles) and the decay (filled triangles). When not shown, the errors are of the same size as the points. As one can see, both colors correlate with flux.

\subsection{Timing analysis}

   \begin{figure}
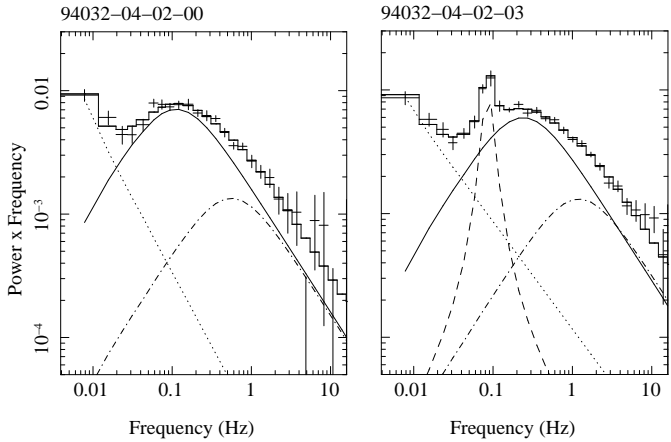

   \begin{tabular}[l]{ll}
   \includegraphics[bb=0 0 588 780,width=4.48cm,clip]{lowfl.eps}   &    \includegraphics[bb=90 0 588 780,width=3.8cm,clip]{highfl.eps} \\
      \end{tabular}
        \caption{\small{Example of low-luminosity ($L_\mathrm{X}=2.31\times10^{36}$ erg s$^{-1}$, left panel) and high-luminosity PSD ($L_\mathrm{X}=2.50\times10^{37}$ erg s$^{-1}$, right panel) with fitted components: a power law (dotted line), $L_{1}$ (solid line), $L_{2}$ (dash-dotted line) and the QPO (dashed line). On the top of each panel the \emph{RXTE} observation ID is reported.}}
      \label{fig:psd}
  \end{figure}

We obtained one average PSD for each observational interval (pointing) as explained above. Our selection of time resolution and length of segments implies a frequency range of $\sim$0.008--16 Hz. Thus, peaks derived from the \  {neutron star's} pulsations do not affect the analysis of the PSD due to their very low frequency, being $\nu_\mathrm{pulse} \sim 2$ mHz the frequency of the fundamental peak.\\

In order to have a unified phenomenological description of the timing features during the outburst, we fitted all the noise components, with the exception of the poorly constrained lower-frequency one, with Lorentzian functions, each one denoted as $L_i$, 
where $i$ determines the type of component. The characteristic frequency 
$\nu_\mathrm{max}$ of $L_i$ is denoted $\nu_i$. This is the frequency where the component contributes most of its variance per logarithmic frequency interval and is defined \citep{bel02} as 

\begin{equation}
\nu_\mathrm{max} = \sqrt{\nu_0^2 + (FWHM/2)^2}~, 
\end{equation}

\noindent
where $\nu_0$ is the centroid frequency and $FWHM$  the full width at half maximum of the Lorentzian component. All the frequencies reported in this work refer to characteristic frequencies $\nu_\mathrm{max}$. To fit the noise below $\sim$0.05 Hz, a power law was employed.\\

\subsubsection{Discovery of a QPO}

A QPO is detected in the PSD of 1A 1118-615 during the brightest observations. When the X-ray luminosity goes below $\sim$40\% of the maximum luminosity, that is for flux below 4$\times10^{-9}$ ergs cm$^{-2}$ s$^{-1}$, the QPO disappears. Figure~\ref{fig:psd} shows two examples of PSD at different luminosity to allow comparison: the QPO is evident in the high-luminosity plot (right panel of Fig.~\ref{fig:psd}), while it is undetectable in the low-luminosity one (left panel of Fig.~\ref{fig:psd}). In the figure, the broad-noise components are also shown.\\

\begin{figure}
   \includegraphics[bb=0 100 400 400,width=9cm,clip]{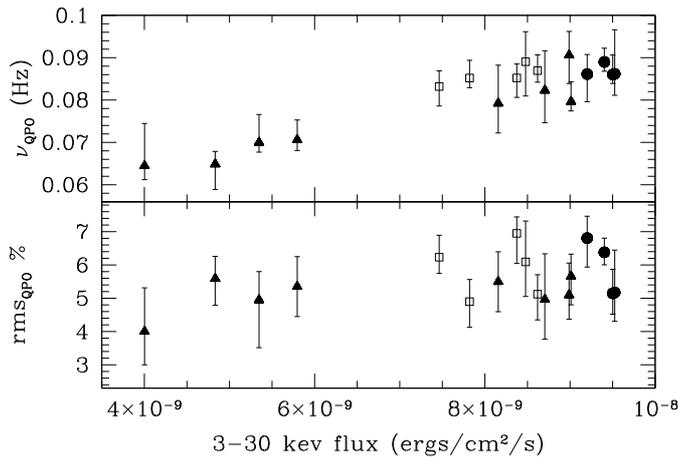}   
  \caption{\small{QPO parameters vs. calculated 3--30 keV flux: characteristic frequency (upper panel) and fractional $rms$ (lower panel). Different symbols mark the different phases of the outburst: rising (empty squares), peak (full circles), and decay (full triangles).}}
      \label{fig:QPO_fl}
   \end{figure}

The QPO was fitted by a narrow Lorentzian. Its maximum frequency, $\nu_\mathrm{QPO}$, is $\sim$0.08 Hz, and the measured quality factor $Q$, obtained as $Q = \nu_0/FWHM$, varies in the range 3--9. Given the relatively large $Q$ value, the centroid frequency $\nu_{0}$ coincides approximately with the maximum frequency $\nu_\mathrm{QPO}$. Figure~\ref{fig:QPO_fl} shows the trend followed by the characteristic frequency and the $rms$ of the QPO feature during the outburst. Both parameters fall within the typical range seen in HMXBs.
The flux shown in Fig.~\ref{fig:QPO_fl} is calculated in the 3--30 keV range from the best-fit spectral model (see Sect.~\ref{sect:spec})\\

   \begin{figure}
   \includegraphics[bb=0 0 400 250,width=9cm,clip]{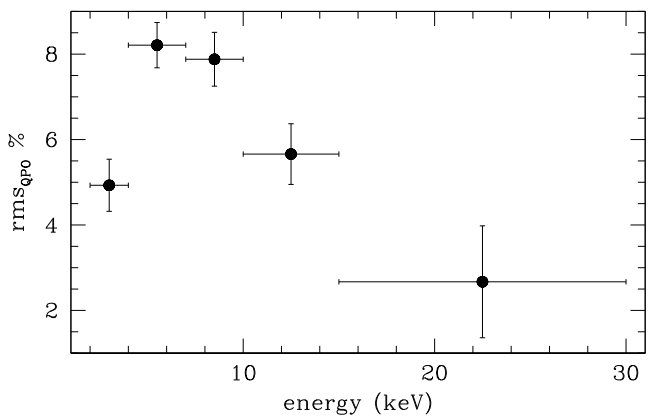}   
  \caption{\small{Fractional $rms$ variability for the 0.08 Hz QPO as a function of energy.}}
      \label{fig:qpo_en}
   \end{figure}

We have also studied the energy dependence of the QPO feature. We extracted PSD in the following energy bands, 2--4, 4--7, 7--10, 10--15, and 15--30 keV, using data from the PCA. We found evidence for a decrease of the QPO variability with energy, for energies  higher than 4 keV. Similar results are obtained repeating the analysis with data from different observations. Results are shown in Fig.~\ref{fig:qpo_en} for observation ID 94032-04-02-02.

\subsubsection{Broad-noise components}

The power spectra continuum was fitted with the sum of, at most, three components. The low-frequency noise ($\lesssim$0.05 Hz) is accounted for by a power law, which is not well constrained. Above 0.05 Hz, the noise is well described by two zero-centered Lorentzians. $L_\mathrm{1}$ dominates below 1 Hz, with a characteristic frequency varying between 0.03--0.28 Hz. When $L_\mathrm{2}$ is not present, it also describes the noise up to 16 Hz. This component, together with the power law, is the only one that is constantly present along the outburst. With the highest fractional $rms$ (up to 28\%), this is the best constrained component. $L_\mathrm{2}$ accounts instead for the high-frequency noise, with $\nu_\mathrm{2} \sim$ 0.5--2.1 Hz. It is not detectable in the last four pointings due to the low signal-to-noise in the corresponding frequency range at low luminosities. Its $rms$ varies between 4\%--14\%.\\

\begin{figure}
   \includegraphics[width=9cm]{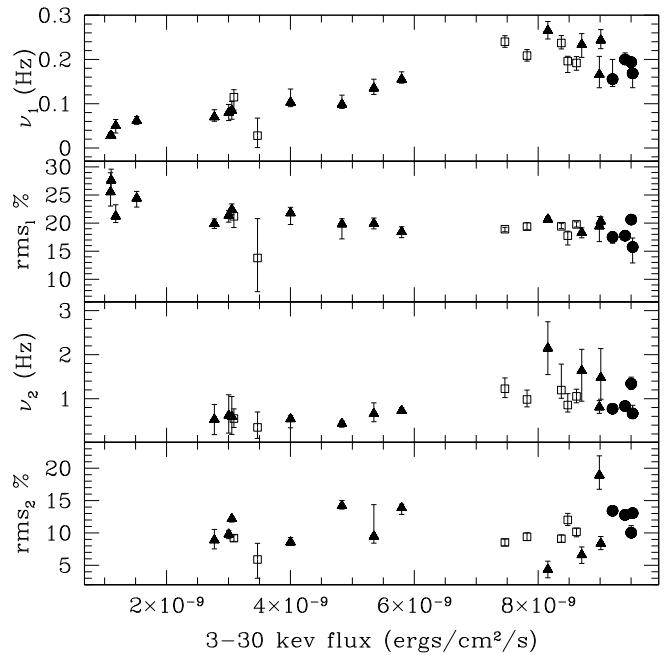}   
  \caption{\small{Frequency and fractional amplitude variability as a function of X-ray flux for the $L_{1}$ (first two panels from above) and $L_{2}$ components (third and fourth panel from above). Different symbols mark the different phases of the outburst: rising (empty squares), peak (full circles), and decay (full triangles).}}
      \label{fig:noise_fl}
   \end{figure}

\begin{figure}
   \includegraphics[bb=0 110 400 400,width=9cm,clip]{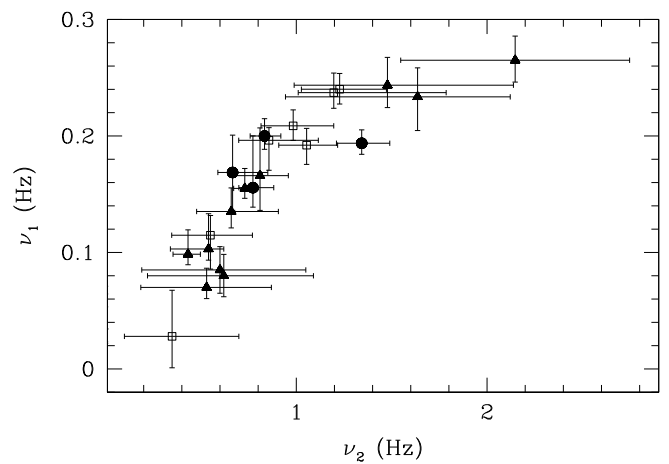}   
  \caption{\small{Relation between the characteristic frequency of the two broad-noise components $L_{1}$ and $L_{2}$. Different symbols mark the different phases of the outburst: rising (empty squares), peak (full circles), and decay (full triangles).}}
      \label{fig:nucompa1}
   \end{figure}

In Fig.~\ref{fig:noise_fl} we present the behavior of the broad-noise components $L_{1}$ and $L_{2}$ as a function of the measured flux, while in Fig.~\ref{fig:nucompa1} the relation between the characteristic frequencies $\nu_{1}$ and $\nu_{2}$ is shown. 
As the flux increases, the characteristic frequency increases and the amplitude of variability decreases. This trend is significant for $L_{1}$, while is much weaker for $L_{2}$. Another interesting feature in  Fig.~\ref{fig:noise_fl} is that above a certain flux the variation of the timing parameters with X-ray flux saturates. Due to the observational gap it is difficult to pin down the exact value of the flux at which this flattening occurs, but it lies in the range 6.5--7.5$\times10^{-9}$ erg cm$^{-2}$ s$^{-1}$.

\subsubsection{X-ray pulsations}

   \begin{figure}
   \includegraphics[bb=15 90 400 400,width=9.5cm,clip]{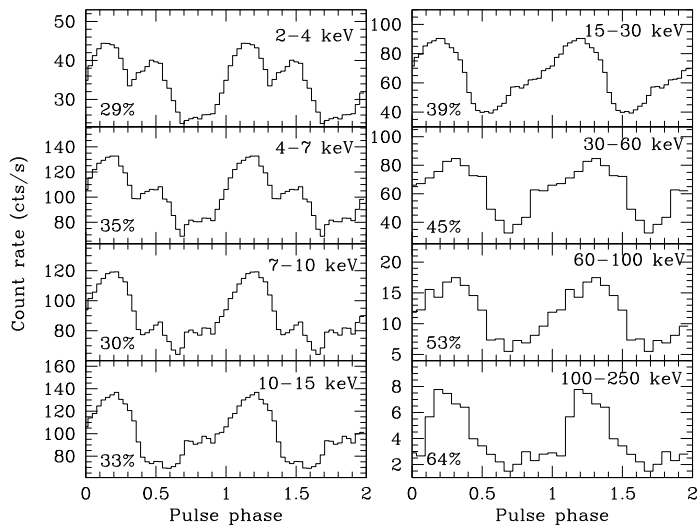}   
  \caption{\small{Folded light curves of 1A 1118-615 in five energy bands. The
folding epoch is MJD 54843. Profiles with energy below 30 keV were obtained
from PCA light curves, while above this energy from HEXTE data. Two
complete cycles are shown for clarity The number in the lower-left corner
represents the pulse fraction. }}
      \label{fig:pulses}
   \end{figure}

The 407-s X-ray pulsation is clearly seen in the light curves of 1A
1118-615 at all energies.  However, the precise
determination of the spin period is hampered by its relatively long value
($\sim$407 s) in comparison with the duration of the observational
intervals. We selected one of the observations with the longest on-source time
(MJD 54843.9750) and derived the spin period by epoch-folding, that is, by
folding data over a period range and obtaining the chi-square of the folded
light curve. The period that gives the highest $\chi^2$ is taken as the
best-fit period. We found 407.5 s. This period was refined by
cross-correlating several pulse profiles obtained by cutting the light
curve into 1230-s segments with a template made up from the
observation-averaged profile. The phase shifts were fitted to a linear
function, whose slope represents the correction in frequency  to the actual
period. A new template was derived for this corrected period, and the
process repeated. The error was estimated from the uncertainty of the slope
from the linear fit. The final spin period was  407.77$\pm$0.08 s. Pulse
profiles were obtained by folding light curves at different energies with
the best-fit period.\\

Figure~\ref{fig:pulses} shows the pulse profile at various energy bands. We find a
significant change of the pulse profile with energy, with the pulse
profile showing a double peak structure below 10 keV. The intensity of the second peak
decreases as the energy increases. Above 10 keV the second peak completely
disappears and the pulse profile becomes nearly sinusoidal. The pulse
fraction, defined as $(I_{\rm max}-I_{\rm min})/(I_{\rm max}+I_{\rm min})$,
increases as the energy increases above 10 keV. Below 10 keV there is a relative minimum at energies 4--7 keV, in good agreement with the results of \citet{dor10}.

 \subsection{Spectral analysis}   \label{sect:spec}

Energy spectra were best fitted with an absorbed power law with exponential high-energy cutoff and a gaussian line profile at $\sim$6.5 keV to account for Fe K$\alpha$ fluorescence. The inclusion of a cyclotron absorption feature with central energy of $\sim$60 keV was a necessary step in order to obtain acceptable fits. The obtained reduced chi-square was $\chi^{2}_\mathrm{red}$ = 0.9--1.6, \  {for 72 degrees of freedom (d.o.f.).}\\

\begin{figure}
   \includegraphics[width=9cm]{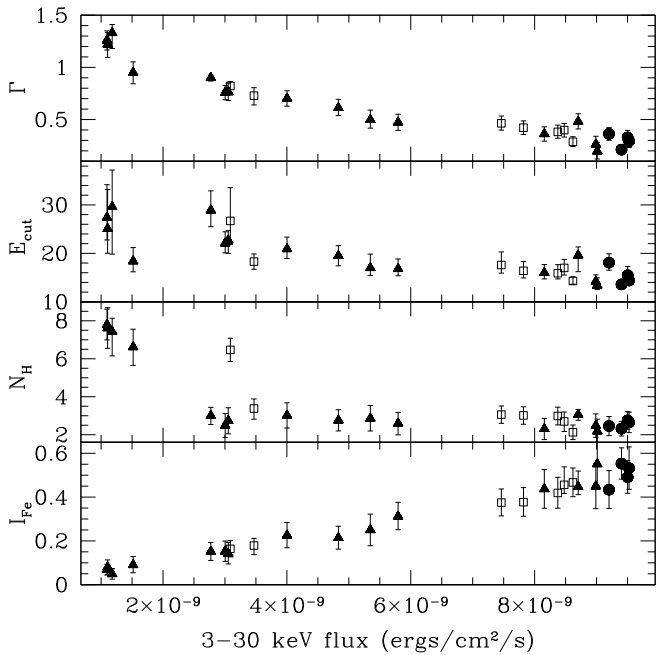}   
  \caption{\small{Main spectral parameters. From above: photon index, cutoff energy (keV), $N_\mathrm{H}$ ($10^{22}$ cm$^{-2}$), Fe line intensity (0.01$\times$photons cm$^{-2}$ s$^{-1}$). Different symbols mark the different phases of the outburst: rising (empty squares), peak (full circles), and decay (full triangles).}}
      \label{fig:spec}
   \end{figure}

Figure~\ref{fig:spec} shows the behavior followed by the main spectral parameters as a function of the flux. Along the outburst, the photon index anti-correlates with flux, implying, as also the colors show, softer emission at low flux and harder emission at high flux. A similar trend is followed by the cutoff energy, with lower values in correspondence to higher flux. The best-fit hydrogen column density, although it can be only poorly constrained by the PCA, remains approximately constant during the outburst evolution at $\sim3 \times 10^{22}$ cm$^{-2}$; only a few observations at the very end of the available data sample and one observation at the beginning show higher values: in these cases, very poor fits were obtained by fixing the value of $N_\mathrm{H}$ at $\sim$3$ \times 10^{22}$ cm$^{-2}$, with $\chi_\mathrm{red}^{2} \sim$2--3 \  {(75 d.o.f.)}. By letting this parameter free to vary, the retrieved $\chi_\mathrm{red}^{2}$ lowered to $\sim$1.05 \  {(76 d.o.f.)}, and the value of $N_\mathrm{H}$ was $\sim$7$\times 10^{22}$ cm$^{-2}$. \\

Using the source infrared excess $E(J - K)=0.47$ \citep{coe94} and assuming the mean extinction law ($R_{V} = 3.1$), we obtained the total measured visual extinction from $A_{V}/E(J - K) = 5.82\pm0.10$ \citep{rieke85}. We converted it to the corresponding hydrogen column density value from $N_\mathrm{H}/A_{V}	= 1.79\pm0.03\times10^{21}$ atoms cm$^{-2}$ mag \citep{pred95} in order to compare the interstellar value of $N_\mathrm{H}$, as obtained from the extinction, with the \  {total one}, provided by X-ray data. We obtained a visual extinction of $A_{V}=2.7$ mag, which corresponds to $N_\mathrm{H}=4.9\times10^{21}$ atoms cm$^{-2}$. This value is one order of magnitude lower than the one obtained by our X-ray data. This difference can be explained assuming that the source of extinction measured by the X-ray band differs from that measured in the optical and infrared bands. The derived optical/IR $N_\mathrm{H}$ measures the interstellar extinction, while the $N_\mathrm{H}$ obtained from the X-ray spectrum is also affected by absorbing material locally, that is, in the vicinity of the compact object. \  {An additional local source of extinction could} be the Be star circumstellar disk, from which the neutron star accretes.\\

The energy of the Fe fluorescence line does not vary. We obtained a weighted mean of 6.45$\pm$0.03 keV. The strength of the Fe gaussian line is strongly correlated with flux, pointing to an increasing fraction of reprocessed material by the circumstellar disk while the flux increases. The gaussian line width was almost constant during the outburst, and was eventually fixed at 0.5 keV in all the fits. The equivalent width (EW) of the line remained fairly constant during the outburst, with a weighted mean value of 0.12$\pm$0.04 keV, with the exception of the few observations in which $N_\mathrm{H}$ reached the highest values. In this case, also the iron line EW increased to higher values, around 0.23 keV. \\

\subsubsection{Pulse-phase spectroscopy}

\begin{figure}
   \includegraphics[width=9cm]{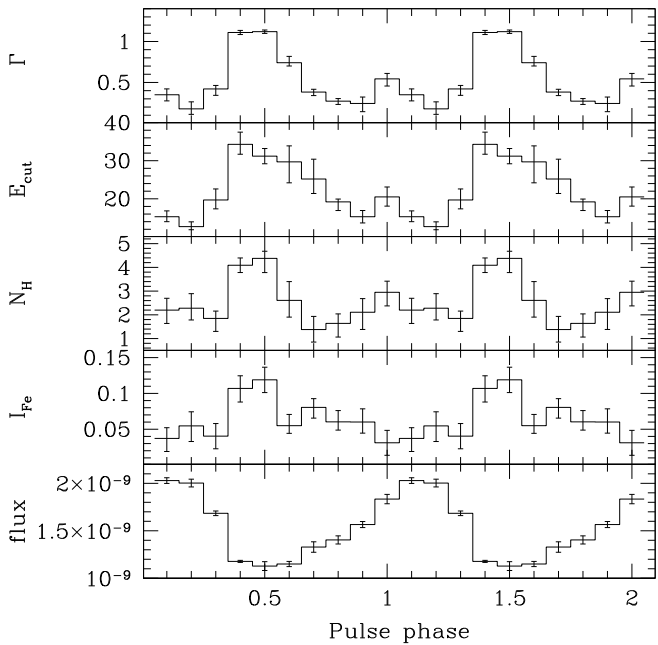}   
  \caption{\small{Pulse-phase dependence of the main spectral parameters and X-ray flux. From top to bottom: photon index, cutoff energy (keV), hydrogen column density ($10^{22}$ cm$^{-2}$), Fe line intensity (0.01$\times$photons cm$^{-2}$ s$^{-1}$), 3--20 keV flux (erg cm$^{-2}$ s$^{-1}$).}}
      \label{fig:phaseSpec}
   \end{figure}

We performed spin phase-resolved spectroscopy for one of the longest pointings. The phase was divided into 10 bins, and for each bin a spectrum was extracted and fitted in the 3--20 keV energy range with the same components as the phase-averaged spectra. The cyclotron absorption feature was not included at this stage, since it lies outside the sensitivity range of the PCA. In the 3-20 keV range, its contribution is negligible. Five of the ten obtained spectra showed residuals at $\sim$8 keV, consistent with an emission-like feature, which was not significant in the phase-averaged spectra. The physical origin of this feature is uncertain \citep{cob02,rod09,dor10}. This was fitted with an additional gaussian.  Results are shown in Fig.~\ref{fig:phaseSpec}.  The modulation of the 3--20 keV flux reflects, as expected, the pulse profiles of Fig.~\ref{fig:pulses}. The spectral parameters clearly exhibit phase dependence: the photon index, the cutoff energy, $N_\mathrm{H}$, and the strength of the iron line show larger values during the off-pulse phase, and lower values in correspondence to the on-pulse. The energy of the iron line did not show significant variation.

\subsubsection{Cyclotron resonant scattering feature}

The cyclotron line was modeled using the GABS XSPEC component, a gaussian absorption line described by the following profile:

\begin{equation}
I(E) = \exp(-(\tau/\sqrt{2\pi}~\sigma) \exp(-0.5((E - E_\mathrm{c}) / \sigma)^2)),
\end{equation}

\noindent where $E_\mathrm{c}$ is the line energy in keV, $\sigma$ is the line width in keV, and $\tau$ is the optical depth at the line center. The obtained central energy of the line was $\sim$60 keV, with some dispersion during the outburst. No harmonics were found in our energy range. This feature was significantly detected during the whole outburst, with the exception of the last four observations, corresponding to an X-ray flux lower than $2\times10^{-9}$ erg cm$^{-2}$ s$^{-1}$. \\

In contrast to other X-ray pulsars, \citep[see, for instance,][]{mih98,tsy06,sta07}, there is no clear correlation between the central energy of the CRSF and the X-ray luminosity, although, on average, $E_\mathrm{c}\sim65$ keV at low luminosities, and $E_\mathrm{c}\sim55$ keV at high luminosities. Errors are quite large for this variation to be statistically significant. Nevertheless, the observed trend of higher energy at low luminosity \  {agrees} with the relation seen in other HMXB pulsars that show an anticorrelation between $E_\mathrm{c}$ and luminosity. The value of $\sim$60--65 keV is the highest energy measured for a cyclotron absorption feature of an X-ray pulsar. We thus confirm and extend to lower luminosities the findings by \citet{dor10}, who found a value of $\sim$55 keV during the peak of the outburst. \\

The presence of CRSFs  provides a tool for direct measurement of the magnetic-field strength of accreting pulsars from the relation

\begin{equation}
E_\mathrm{c} = 11.6~ B_{12} \times (1+z)^{-1}~~ \mathrm{keV}, 
\end{equation}

\noindent where $B_{12}$ is the magnetic field strength in units of 10$^{12}$ G and $z$ is the gravitational redshift in the line-forming region \citep{was83}.
Assuming a gravitational redshift of $z = 0.3$ for a typical neutron-star mass of 1.4 $M_{\sun}$ and radius of 10 km, we estimated the magnetic field to be 6.7$\times 10^{12}$ G. Due to the high-energy CRSF, the value obtained for the magnetic field is among the highest measured for an X-ray pulsar.

\subsection{Correlations between spectral and timing characteristics}

   \begin{figure}
   \includegraphics[bb=0 120 400 400,width=9cm,clip]{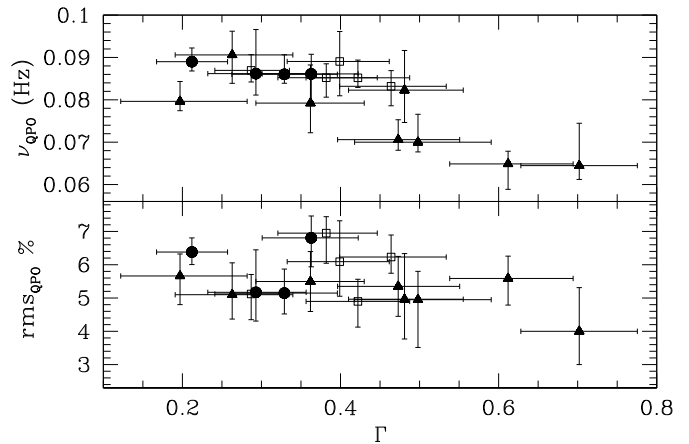}   
  \caption{\small{Relation between the photon index and the QPO characteristic frequency (upper panel) and $rms$ (lower panel). Different symbols mark the different phases of the outburst: rising (empty squares), peak (full circles), and decay (full triangles).}}
      \label{fig:gammaQPO}
   \end{figure}

   \begin{figure}
   \includegraphics[bb=0 120 400 400,width=9cm,clip]{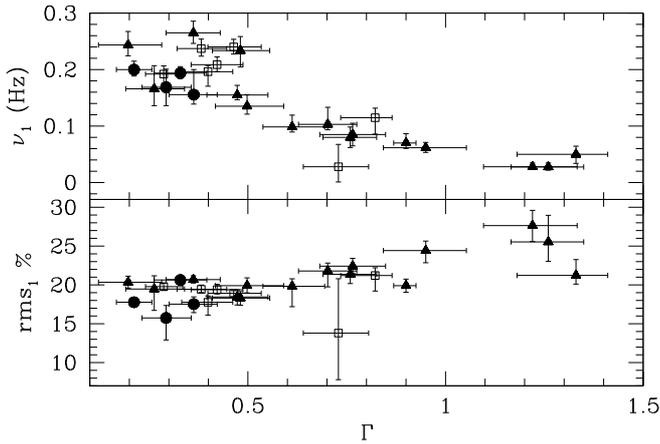}   
  \caption{\small{Same as Fig.~\ref{fig:gammaQPO} for the $L_{1}$ broad-noise component. Different symbols mark the different phases of the outburst: rising (empty squares), peak (full circles), and decay (full triangles).}}
      \label{fig:gamma1}
   \end{figure}

Figures~\ref{fig:gammaQPO} and \ref{fig:gamma1} show the relation between the photon index and the parameters of the QPO and the $L_{1}$ component, respectively. The frequency of both components shows a negative correlation with the photon index; the $L_{1}$ $rms$ shows a correlated behavior with $\Gamma$, while $rms$$_\mathrm{QPO}$ remains constant. The stronger correlation seen for $L_{1}$ can be attributed to a wider range of variability of $\Gamma$, which, in turn, results from a wider range of variability in the X-ray flux.


\section{Discussion}
We have performed detailed spectral and timing analysis of the accreting pulsar 1A~1118-615 during an entire giant outburst.\\

\subsection{Aperiodic variability analysis}

We reported the discovery of a QPO in the power spectra of
the system {\ {(Fig.~\ref{fig:psd})}}. This peaked component has a characteristic
frequency of 0.08 Hz, and it is only observable in high-flux
observations, down to  $L_\mathrm{X}\sim40\%$ of the maximum observed 
luminosity. Including 1A 1118-615, QPOs
have been detected in 17 accretion-powered high-magnetic field
pulsars, comprising many HMXBs and a few LMXBs, both
transient and persistent \citep[see][and references therein]{mar10}. QPOs in X-ray pulsars provide strong evidence for
the presence of an accretion disk during giant outbursts because
they are believed to be related to inhomogeneities in
the inner disk \citep{pau98}. While black-hole binaries and LMXBs show
QPOs with frequencies from a few Hz to a few hundred Hz,
high-magnetic field neutron stars only show low-frequency
QPOs, in the range between 10 mHz and about 1 Hz.\\

The presence of QPOs in the power spectral density of X-ray pulsars is in general explained with either the Magnetospheric Beat Frequency Model \citep[BFM,][]{alp85} or the Keplerian Frequency Model \citep[KFM,][]{van87a}. The main difference between the two models is the interpretation of the QPO. In the BFM, $\nu_\mathrm{QPO}=\nu_\mathrm{k} - \nu_{s}$, where $\nu_\mathrm{k}$ is the Keplerian frequency of the inner accretion disk, and $\nu_\mathrm{s}$ is the spin frequency, while in the KFM, $\nu_\mathrm{QPO} = \nu_\mathrm{k}$. Given that in 1A 118-615 $\nu_\mathrm{s} \ll \nu_\mathrm{QPO}$, it is very difficult to distinguish between the two models. A third model, the magnetic disk precession model \citep{shi02} attributes mHz QPOs to warping/precession modes induced by magnetic torques near the inner edge of the accretion disk. None of the three models can explain the origin of QPOs in all the systems.\\

If one  assumes that QPOs are produced as a result of Keplerian motion of inhomogeneities in an accretion disk, then the longer implied timescales of the QPOs in HMXBs are somehow expected since the inner radius of the accretion disk must be larger than the magnetospheric radius $R_{M}\sim10^{8}$ cm. The inner disk radius can be expressed as $r_{0} = (GM / 4\pi^{2} \nu_\mathrm{k}^{2})^{1/3}$, where $\nu_\mathrm{k}$ is the Keplerian rotation frequency at the inner edge.  An 80 mHz QPO implies a disk radius of 8--9$\times10^{8}$ cm, that is, outside the magnetosphere, as models foresee. \\

The QPO frequency can be employed to estimate the neutron-star magnetic field. The Alfv\'en radius for a 1.4 $M_{\sun}$ neutron star with radius of 10 km can be approximated to $R_{M} \simeq 3\times10^{8}L_{37}^{-2/7}\mu_{30}^{4/7}$ cm  \citep{fra02}, where $L_{37}$ is the X-ray luminosity in units of $10^{37}$ erg s$^{-1}$, and $\mu_{30}$ is the magnetic moment in units of 10$^{30}$ G cm$^{3}$. Assuming that the QPO is formed at the inner accretion disk, that is, equating the inner accretion disk radius as obtained from the QPO frequency, $r_{0}$, with $r_{M}$, we estimated a magnetic moment of 7.8 $\times10^{30}$ G cm$^{3}$, which is equivalent to a magnetic field of 7.8 $\times10^{12}$ G.\\

A positive correlation is expected between the QPO frequency center and the X-ray flux, because at larger mass accretion rate, corresponding to increasing flux, the size of the magnetosphere decreases, and the accretion disk is expected to extend closer to the neutron star. This implies shorter characteristic time-scales, which means higher frequencies. In the case of 1A~1118-615 during the outburst analyzed in this work, we have indeed detected a positive correlation between the QPO frequency and the 3--30 keV flux \ {(Fig.~3)}. In contrast, the fractional $rms$ amplitude does not show any correlation with flux. Note that the overall $rms$ of the continuum remained fairly constant during the outburst. Thus, the disappearance of the QPO cannot be linked to any overall increase in power, but must be attributed to a decrease in the strength of the QPO itself. \\

 We found a negative correlation with energy above 4 keV \ {(Fig. 4)}. No general trend has been identified for QPOs in HMXBs, with cases of correlation \citep[as V0332+53, see][]{qu05}, anti-correlation \citep[XTE J1858+034, KS 1947+300, see][]{muk06,mar10}, or no dependence \citep[EXO 2030+375, see][]{ang89}.  \\

\ {The characteristic frequencies of the two broad-band noise components increase as the flux increases \ {(Fig.~5)}}. This is particularly true for $L_{1}$, the best constrained and always present component. The correlated behavior is at least detected up to flux $\sim$6$\times10^{-9}$ erg cm$^{-2}$ s$^{-1}$ ($L_\mathrm{X}=1.8\times10^{37}$ erg s$^{-1}$, for a distance of 5 kpc). If one assumes that all the aperiodic phenomenology arises in the same region, the correlation observed between frequencies and X-ray flux is somehow expected, and can be explained in the same manner as for the QPO frequency, namely, due to the shrinking of the magnetosphere as flux increases. \\

\ {The fractional amplitude of variability of the $L_1$ component decreases as the flux increases (Fig. 5). This trend was not seen in the QPO $rms$ due to the limited flux range in which the QPO was detected. This is also true for the $L_2$ component, although in this case other effects may be at work. Changes in the mass accretion rate will affect first large radii, i.e., low frequencies. Large amplitude variations may then be suppressed as they propagate inward. For fluxes larger than $\sim7\times10^{-9}$ erg cm$^{-2}$ s$^{-1}$, the correlation of frequencies and variability amplitudes with flux is no more observed. We propose that this saturation corresponds to the minimum extent of the magnetosphere, or, equivalently, to the minimum inner accretion disk radius.}\\

\subsection{Spectral analysis}

 The most remarkable feature in the X-ray energy spectra of 1A~1118-615 is a cyclotron line at $\sim$60 keV, the highest value measured in a HMXB.  This value is consistent with recent results by \citet{dor10}. Before this discovery, the X-ray pulsar showing the highest CRSF energy was A0535+26, with a value of 52 keV \citep{pos08}. It has been pointed out that the energies of cyclotron absorption features vary with X-ray luminosity \citep{mih04,nak06}. Such variations are interpreted in terms of the height of the line-forming region above the neutron star surface. Both positive and negative correlations between the cyclotron line energy and X-ray luminosity were observed in X-ray pulsars, with a prevalent anti-correlation in high-luminosity pulsars and correlation in low-luminosity ones. Prominent examples for such a variability are V0332+53 and 4U 0115+63 on one side, where the cyclotron line energy decreases with luminosity \citep{tsy06,nak06}, and Her X-1 on the other side, where the line energy increases with luminosity \citep{sta07}. A counterexample is on the other hand represented by A0535+262, where the line energy does not show any kind of correlation with flux \citep{cab07}. In the case of 1A~1118-615, no strict trend associated with flux was detected, although higher values for the central energy were measured at lower luminosity. The properties of cyclotron lines are directly related to the pulsar X-ray emission processes and the behavior of matter in strong magnetic fields \citep{mes85}. Therefore, CRSFs carry a wealth of information about the environment in which they are formed. CRSFs provide the only direct measurement of the neutron star magnetic field, that we calculated to be $\sim$6.7$\times10^{12}$ G.  Given the order-of-magnitude calculation, this is compatible with the value of 7.8$\times10^{12}$ G, obtained independently from timing analysys.\\
 
The photon index anti-correlates with flux {\ {(Fig.~{\ref{fig:spec})}}, in contrast to what was found in other Be/XRBs \citep[like, for instance, EXO 2030+375, V0332+53, see][]{wil08,mow06}, but similar to what was observed in SAX~J2103.5+4545 \citep{rei10}. This trend implies a hardening of the spectrum as the flux increases. The hard emission from X-ray pulsars is believed to originate in the accretion column, as a result of Comptonization of soft photons coming from the neutron star thermal mound by high-energy electrons in the accretion flow \citep{bec07}. We speculate that as the accretion rate increases, the photons spend more time, on average, being upscattered in the flow before escaping from the accretion column, thus resulting in harder X-ray spectra \citep{lan82,bec07}. \\

The measured hydrogen column density was almost constant ($N_\mathrm{H}$ at $\sim$3$ \times 10^{22}$ cm$^{-2}$) with time, with the exception of a few observations at the beginning and at the and of the outburst, where it more than doubled its value ($N_\mathrm{H}$ at $\sim$7$ \times 10^{22}$ cm$^{-2}$). After obtaining very poor fits by constraining the value of $N_\mathrm{H}$ to 3$ \times 10^{22}$ cm$^{-2}$ for all pointings, we believe that the observed variation is a real phenomenon. We measured the duration of this ``$N_\mathrm{H}$ transition'', from lower to higher values, obtaining $\sim$3 days in both cases, at the beginning and at the end of the outburst. The time interval with $N_\mathrm{H}$ confined at lower values is approximately 15 days, between the two transitions. This phenomenon is a remarkable spectral feature of 1A~1118-615 and must arises from some change in the matter surrounding the neutron star during the outburst, possibly due to the Be companion activity. It is very unlikely that the observed 15 days interval has an orbital origin since, according to the pulse period/orbital period diagram \citep{cor86}, a 407 s $P_\mathrm{s}$ would imply a 400--800 days $P_\mathrm{orb}$. Also, the lack of Type-I and scarcity of Type-II outbursts seen in this system would contrast with such a short orbital period \citep[see][]{oka01}. More measurements by observatories with high sensitivity at low energies are needed to investigate this peculiar property.  \ {The hydrogen column density as obtained from our data was compared with the value converted from the visual extinction $A_{V}$. A difference of one order of magnitude points to a source of absorption that is local to the compact object, consistently with the on-going X-ray giant outburst.}\\

The center of the Fe K$\alpha$-emission line was determined to be 6.45$\pm$0.03 keV during the entire outburst. This value is compatible with neutral or weakly ionized iron. 
The line equivalent width is approximately constant with flux, while the strength of the line grows linearly with flux {\ {(Fig.~{\ref{fig:spec})}}. This result gives firm evidence that the iron line in fact arises from the fluorescence of continuum X-rays absorbed by cool matter surrounding the central source. High hydrogen column density and the presence of the Fe line during all the outburst attest to a large amount of absorbing material around the compact object. This is a further evidence of the extremely dense and extended disk around the Be companion, which is the Be star displaying the strongest Balmer emission lines \citep{mot88}. \\

\subsection{Pulse profile analysis}

We also studied the pulse profiles at different energies, finding a dependence of the pulse shape with the energy band {\ {(Fig.~{\ref{fig:pulses})}}. A double peak structure is observed below $\sim$10 keV, while a single peak profile is detected above that energy. The pulsed fraction increases with energy from $\sim$30\% to $\sim$65\%, for energy above $\sim$10 keV, with a ``plateau'' at lower energies, \ {in agreement with \citet{dor10}}, similar to what was found in 4U 0115+63 \citep{fer09}. Various types of pulse profiles are observed in X-ray pulsars. The profile energy dependence displayed by 1A~1118-615 is of the same kind as, for instance, 4U 0115+63 \citep{tsy07} and Her X-1 \citep{nag89}. The presence of a second, weaker, peak is usually interpreted as the partially obscured emission of the column of the antipodal spot \citep{tsy07}. Pulse profile changes may have various origins, like changes in the spot shape and position, variations of the emission pattern, of the inner disk radius or the optical depth of the accretion stream.\\

\ { To investigate the dependence of the main spectral parameters with the spin phase, we performed spin phase-resolved spectroscopy (Fig.~{\ref{fig:phaseSpec})}.} Our results show a clear modulation of spectral parameters with the spin phase. In particular, spectra are softer at off-pulse phases, and harder at on-pulse phases. The cutoff energy, the hydrogen column density and the intensity of the iron fluorescence line display the higher values at the off-pulse phase. We speculate that a hardening of the spectrum as the spin phase gets on-pulse may be due to the fact that the accretion column, and thus the Comptonized X-ray emission, are being observed from a more direct and somehow transparent line of sight. On the other hand, at the off-pulse phase, much more material interpose between the emission flow and the observer, resulting in softer spectra, higher hydrogen column density, and larger quantity of reprocessed iron, as the spectral parameters coherently indicate. \\

\subsection{Spectral/timing correlations}
\ {Figure 11 shows the relationship between the photon index and the frequency and $rms$ variability of the $L_1$ noise component. This relationship results naturally from the correlations reported above of the photon index and the timing parameters with flux. However, Fig.~11 allows a quick comparison of the X-ray temporal variability as a function of spectral shape. This is the first time that such correlation is reported in an accretion-powered pulsar. Figure 10 shows the same relationship for the QPO. The correlation is much weaker than for $L_1$, possibly due to the limited flux range where the QPO is observed.}\\

Figure 11 indicates that the source is more variable when it shows a soft spectrum, \emph{i.e.}, at low flux and, more importantly, that the physical region where the aperiodic variability originates and the emission region are strongly linked to each other. The aperiodic variability is supposed to arise from a region outside the magnetosphere at $R\gtrsim 10^8$ cm, while the emitting region that originates the energy spectra lies in the accretion column, i.e., at distances close to the neutron star surface, namely $10^6-10^7$ cm. Nevertheless, these two regions are somehow physically connected. The correlation between $rms$ and photon index might be explained by simple Comptonization models, if one assumes that the source of variability is variations in the soft photon input. In this case, low-energy photons would retain most of the variability of the seed photon input since they did not spend much time in the Comptonization medium and have not undergone many scatterings. In contrast, the variability of high-energy photons would be smeared out since these photons have spent longer time in the accretion flow. The overall result would be higher variability in correspondence to a soft spectrum, as observed.} \\  
  
This work demonstrates the need for a comprehensive approach to the study of accretion-powered pulsars. The complexity of the models of accretion in these systems, mainly due to the strong coupling of the magnetic field and the accretion flow, crucially requires the employment of all kinds of techniques, multiband photometry (colors), spectral and timing analysis.
    \begin{acknowledgements}
The work of EN is partly supported by the Spanish Ministerio de Educaci\'on y Ciencia, and FEDER, under contract AYA 2007-62487. EN acknowledges a ``V Segles'' research grant from the University of Valencia. This work has been supported in part by the European Union Marie Curie grant MTKD-CT-2006-039965 and EU FP7 ``Capacities'' GA No206469.
      \end{acknowledgements}

\bibliographystyle{../aa}

\begin{thebibliography}{50}
\expandafter\ifx\csname natexlab\endcsname\relax\def\natexlab#1{#1}\fi

\bibitem[{{Alpar} \& {Shaham}(1985)}]{alp85}
{Alpar}, M.~A. \& {Shaham}, J. 1985, \nat, 316, 239

\bibitem[{{Angelini} {et~al.}(1989){Angelini}, {Stella}, \& {Parmar}}]{ang89}
{Angelini}, L., {Stella}, L., \& {Parmar}, A.~N. 1989, \apj, 346, 906

\bibitem[{{Arnaud}(1996)}]{arn96}
{Arnaud}, K.~A. 1996, in Astronomical Society of the Pacific Conference Series,
  Vol. 101, Astronomical Data Analysis Software and Systems V, ed. G.~H.
  {Jacoby} \& J.~{Barnes}, 17--+

\bibitem[{{Becker} \& {Wolff}(2007)}]{bec07}
{Becker}, P.~A. \& {Wolff}, M.~T. 2007, \apj, 654, 435

\bibitem[{{Belloni} \& {Hasinger}(1990)}]{bel90}
{Belloni}, T. \& {Hasinger}, G. 1990, \aap, 230, 103

\bibitem[{{Belloni} {et~al.}(2002){Belloni}, {Psaltis}, \& {van der
  Klis}}]{bel02}
{Belloni}, T., {Psaltis}, D., \& {van der Klis}, M. 2002, \apj, 572, 392

\bibitem[{{Bradt} {et~al.}(1993){Bradt}, {Rothschild}, \& {Swank}}]{bra93}
{Bradt}, H.~V., {Rothschild}, R.~E., \& {Swank}, J.~H. 1993, \aaps, 97, 355

\bibitem[{{Caballero} {et~al.}(2007){Caballero}, {Kretschmar}, {Santangelo},
  {Staubert}, {Klochkov}, {Camero}, {Ferrigno}, {Finger}, {Kreykenbohm},
  {McBride}, {Pottschmidt}, {Rothschild}, {Sch{\"o}nherr}, {Segreto}, {Suchy},
  {Wilms}, \& {Wilson}}]{cab07}
{Caballero}, I., {Kretschmar}, P., {Santangelo}, A., {et~al.} 2007, \aap, 465,
  L21

\bibitem[{{Coburn} {et~al.}(2002){Coburn}, {Heindl}, {Rothschild}, {Gruber},
  {Kreykenbohm}, {Wilms}, {Kretschmar}, \& {Staubert}}]{cob02}
{Coburn}, W., {Heindl}, W.~A., {Rothschild}, R.~E., {et~al.} 2002, \apj, 580,
  394

\bibitem[{{Coe} {et~al.}(1994){Coe}, {Roche}, {Everall}, {Fishman}, {Hagedon},
  {Finger}, {Wilson}, {Buckley}, {Shrader}, {Fabregat}, {Polcaro},
  {Giovannelli}, \& {Villada}}]{coe94}
{Coe}, M.~J., {Roche}, P., {Everall}, C., {et~al.} 1994, \aap, 289, 784

\bibitem[{{Corbet}(1986)}]{cor86}
{Corbet}, R.~H.~D. 1986, \mnras, 220, 1047

\bibitem[{{Doroshenko} {et~al.}(2010){Doroshenko}, {Suchy}, {Santangelo},
  {Staubert}, {Kreykenbohm}, {Rothschild}, {Pottschmidt}, \& {Wilms}}]{dor10}
{Doroshenko}, V., {Suchy}, S., {Santangelo}, A., {et~al.} 2010, \aap, 515, L1+

\bibitem[{{Eyles} {et~al.}(1975){Eyles}, {Skinner}, {Willmore}, \&
  {Rosenberg}}]{eyl75}
{Eyles}, C.~J., {Skinner}, G.~K., {Willmore}, A.~P., \& {Rosenberg}, F.~D.
  1975, \nat, 254, 577

\bibitem[{{Ferrigno} {et~al.}(2009){Ferrigno}, {Becker}, {Segreto}, {Mineo}, \&
  {Santangelo}}]{fer09}
{Ferrigno}, C., {Becker}, P.~A., {Segreto}, A., {Mineo}, T., \& {Santangelo},
  A. 2009, \aap, 498, 825

\bibitem[{{Frank} {et~al.}(2002){Frank}, {King}, \& {Raine}}]{fra02}
{Frank}, J., {King}, A., \& {Raine}, D.~J. 2002, {Accretion Power in
  Astrophysics: Third Edition} (Cambridge University Press)

\bibitem[{{Ives} {et~al.}(1975){Ives}, {Sanford}, \& {Bell Burnell}}]{ive75}
{Ives}, J.~C., {Sanford}, P.~W., \& {Bell Burnell}, S.~J. 1975, \nat, 254, 578

\bibitem[{{James} {et~al.}(2010){James}, {Paul}, {Devasia}, \&
  {Indulekha}}]{mar10}
{James}, M., {Paul}, B., {Devasia}, J., \& {Indulekha}, K. 2010, \mnras, 407,
  285

\bibitem[{{Janot-Pacheco} {et~al.}(1981){Janot-Pacheco}, {Ilovaisky}, \&
  {Chevalier}}]{jan81}
{Janot-Pacheco}, E., {Ilovaisky}, S.~A., \& {Chevalier}, C. 1981, \aap, 99, 274

\bibitem[{{Langer} \& {Rappaport}(1982)}]{lan82}
{Langer}, S.~H. \& {Rappaport}, S. 1982, \apj, 257, 733

\bibitem[{{Lin} {et~al.}(2010){Lin}, {Takata}, {Kong}, \& {Hwang}}]{lin10}
{Lin}, L., {Takata}, J., {Kong}, A.~K.~H., \& {Hwang}, C. 2010, \mnras, 1586

\bibitem[{{Mereghetti} {et~al.}(1987){Mereghetti}, {Bignami}, {Caraveo}, \&
  {Goldwurm}}]{mer87}
{Mereghetti}, S., {Bignami}, G.~F., {Caraveo}, P.~A., \& {Goldwurm}, A. 1987,
  \apj, 312, 755

\bibitem[{{Meszaros} \& {Nagel}(1985)}]{mes85}
{Meszaros}, P. \& {Nagel}, W. 1985, \apj, 298, 147

\bibitem[{{Mihara} {et~al.}(1998){Mihara}, {Makishima}, \& {Nagase}}]{mih98}
{Mihara}, T., {Makishima}, K., \& {Nagase}, F. 1998, Advances in Space
  Research, 22, 987

\bibitem[{{Mihara} {et~al.}(2004){Mihara}, {Makishima}, \& {Nagase}}]{mih04}
{Mihara}, T., {Makishima}, K., \& {Nagase}, F. 2004, \apj, 610, 390

\bibitem[{{Miyamoto} {et~al.}(1991){Miyamoto}, {Kimura}, {Kitamoto}, {Dotani},
  \& {Ebisawa}}]{miy91}
{Miyamoto}, S., {Kimura}, K., {Kitamoto}, S., {Dotani}, T., \& {Ebisawa}, K.
  1991, \apj, 383, 784

\bibitem[{{Motch} {et~al.}(1988){Motch}, {Pakull}, {Janot-Pacheco}, \&
  {Mouchet}}]{mot88}
{Motch}, C., {Pakull}, M.~W., {Janot-Pacheco}, E., \& {Mouchet}, M. 1988, \aap,
  201, 63

\bibitem[{{Mowlavi} {et~al.}(2006){Mowlavi}, {Kreykenbohm}, {Shaw},
  {Pottschmidt}, {Wilms}, {Rodriguez}, {Produit}, {Soldi}, {Larsson}, \&
  {Dubath}}]{mow06}
{Mowlavi}, N., {Kreykenbohm}, I., {Shaw}, S.~E., {et~al.} 2006, \aap, 451, 187

\bibitem[{{Mukherjee} {et~al.}(2006){Mukherjee}, {Bapna}, {Raichur}, {Paul}, \&
  {Jaaffrey}}]{muk06}
{Mukherjee}, U., {Bapna}, S., {Raichur}, H., {Paul}, B., \& {Jaaffrey},
  S.~N.~A. 2006, Journal of Astrophysics and Astronomy, 27, 25

\bibitem[{{Nagase}(1989)}]{nag89}
{Nagase}, F. 1989, \pasj, 41, 1

\bibitem[{{Nakajima} {et~al.}(2006){Nakajima}, {Mihara}, {Makishima}, \&
  {Niko}}]{nak06}
{Nakajima}, M., {Mihara}, T., {Makishima}, K., \& {Niko}, H. 2006, Advances in
  Space Research, 38, 2756

\bibitem[{{Negueruela}(2007)}]{neg07}
{Negueruela}, I. 2007, in Astronomical Society of the Pacific Conference
  Series, Vol. 367, Massive Stars in Interactive Binaries, ed. {N.~St.-Louis \&
  A.~F.~J.~Moffat}, 477--+

\bibitem[{{Negueruela} {et~al.}(2000){Negueruela}, {Reig}, {Finger}, \&
  {Roche}}]{neg00}
{Negueruela}, I., {Reig}, P., {Finger}, M.~H., \& {Roche}, P. 2000, \aap, 356,
  1003

\bibitem[{{Nowak} {et~al.}(1999){Nowak}, {Vaughan}, {Wilms}, {Dove}, \&
  {Begelman}}]{now99}
{Nowak}, M.~A., {Vaughan}, B.~A., {Wilms}, J., {Dove}, J.~B., \& {Begelman},
  M.~C. 1999, \apj, 510, 874

\bibitem[{{Okazaki} \& {Negueruela}(2001)}]{oka01}
{Okazaki}, A.~T. \& {Negueruela}, I. 2001, \aap, 377, 161

\bibitem[{{Paul} \& {Rao}(1998)}]{pau98}
{Paul}, B. \& {Rao}, A.~R. 1998, \aap, 337, 815

\bibitem[{{Postnov} {et~al.}(2008){Postnov}, {Staubert}, {Santangelo},
  {Klochkov}, {Kretschmar}, \& {Caballero}}]{pos08}
{Postnov}, K., {Staubert}, R., {Santangelo}, A., {et~al.} 2008, \aap, 480, L21

\bibitem[{{Predehl} \& {Schmitt}(1995)}]{pred95}
{Predehl}, P. \& {Schmitt}, J.~H.~M.~M. 1995, \aap, 293, 889

\bibitem[{{Qu} {et~al.}(2005){Qu}, {Zhang}, {Song}, \& {Falanga}}]{qu05}
{Qu}, J.~L., {Zhang}, S., {Song}, L.~M., \& {Falanga}, M. 2005, \apjl, 629, L33

\bibitem[{{Reig} {et~al.}(2010){Reig}, {S{\l}owikowska}, {Zezas}, \&
  {Blay}}]{rei10}
{Reig}, P., {S{\l}owikowska}, A., {Zezas}, A., \& {Blay}, P. 2010, \mnras, 401,
  55

\bibitem[{{Rieke} \& {Lebofsky}(1985)}]{rieke85}
{Rieke}, G.~H. \& {Lebofsky}, M.~J. 1985, \apj, 288, 618

\bibitem[{{Rodes-Roca} {et~al.}(2009){Rodes-Roca}, {Torrej{\'o}n},
  {Kreykenbohm}, {Mart{\'{\i}}nez N{\'u}{\~n}ez}, {Camero-Arranz}, \&
  {Bernab{\'e}u}}]{rod09}
{Rodes-Roca}, J.~J., {Torrej{\'o}n}, J.~M., {Kreykenbohm}, I., {et~al.} 2009,
  \aap, 508, 395

\bibitem[{{Rutledge} {et~al.}(2007){Rutledge}, {Bildsten}, {Brown},
  {Chakrabarty}, {Pavlov}, \& {Zavlin}}]{rut07}
{Rutledge}, R.~E., {Bildsten}, L., {Brown}, E.~F., {et~al.} 2007, \apj, 658,
  514

\bibitem[{{Shirakawa} \& {Lai}(2002)}]{shi02}
{Shirakawa}, A. \& {Lai}, D. 2002, \apj, 565, 1134

\bibitem[{{Staubert} {et~al.}(2007){Staubert}, {Shakura}, {Postnov}, {Wilms},
  {Rothschild}, {Coburn}, {Rodina}, \& {Klochkov}}]{sta07}
{Staubert}, R., {Shakura}, N.~I., {Postnov}, K., {et~al.} 2007, \aap, 465, L25

\bibitem[{{Tsygankov} {et~al.}(2006){Tsygankov}, {Lutovinov}, {Churazov}, \&
  {Sunyaev}}]{tsy06}
{Tsygankov}, S.~S., {Lutovinov}, A.~A., {Churazov}, E.~M., \& {Sunyaev}, R.~A.
  2006, \mnras, 371, 19

\bibitem[{{Tsygankov} {et~al.}(2007){Tsygankov}, {Lutovinov}, {Churazov}, \&
  {Sunyaev}}]{tsy07}
{Tsygankov}, S.~S., {Lutovinov}, A.~A., {Churazov}, E.~M., \& {Sunyaev}, R.~A.
  2007, Astronomy Letters, 33, 368

\bibitem[{{van der Klis} {et~al.}(1987){van der Klis}, {Stella}, {White},
  {Jansen}, \& {Parmar}}]{van87a}
{van der Klis}, M., {Stella}, L., {White}, N., {Jansen}, F., \& {Parmar}, A.~N.
  1987, \apj, 316, 411

\bibitem[{{Villada} {et~al.}(1999){Villada}, {Rossi}, {Polcaro}, \&
  {Giovannelli}}]{vil99}
{Villada}, M., {Rossi}, C., {Polcaro}, V.~F., \& {Giovannelli}, F. 1999, \aap,
  344, 277

\bibitem[{{Wasserman} \& {Shapiro}(1983)}]{was83}
{Wasserman}, I. \& {Shapiro}, S.~L. 1983, \apj, 265, 1036

\bibitem[{{Wilson} {et~al.}(2008){Wilson}, {Finger}, \&
  {Camero-Arranz}}]{wil08}
{Wilson}, C.~A., {Finger}, M.~H., \& {Camero-Arranz}, A. 2008, \apj, 678, 1263

\end{thebibliography}

\end{document}